\documentclass[superscriptaddress,twocolumn,secnumarabic,
 amssymb,amsmath,nobibnotes,aps,prd,showkeys,showpacs,nofootinbib]{revtex4}

\usepackage{graphicx}
\usepackage{epsf}
\usepackage{bm}%
\newcommand{\be}{\begin{equation}}
\newcommand{\ee}{\end{equation}}

\newcommand{\rmm}{\rho_{m}}

\newcommand{\mincir}{\raise
-3.truept\hbox{\rlap{\hbox{$\sim$}}\raise4.truept\hbox{$<$}\ }}
\newcommand{\magcir}{\raise
-3.truept\hbox{\rlap{\hbox{$\sim$}}\raise4.truept\hbox{$>$}\ }}

\begin{document}
\title{Constraints on Cold Dark Matter Accelerating Cosmologies and Cluster Formation}

\author{S. Basilakos}
\affiliation{Academy of Athens, Research Center for Astronomy and Applied Mathematics\\
 Soranou Efesiou 4, 11527, Athens, Greece}

\author{J. A. S. Lima}

\affiliation{Departamento de Astronomia (IAGUSP),  Universidade de S\~ao Paulo\\
Rua do Mat\~ao, 1226, 05508-900, S. Paulo, Brazil}

\begin{abstract}
We discuss the properties of homogeneous and isotropic flat
cosmologies in which the present accelerating stage is powered only
by the gravitationally induced creation of cold dark matter (CCDM)
particles ($\Omega_{m}=1$). For some matter creation rates
proposed in the literature, we show that the main cosmological
functions such as the scale factor of the universe, the Hubble
expansion rate, the growth factor and the cluster formation rate are
analytically defined. The best CCDM scenario has only one free
parameter and our joint analysis involving BAO + CMB + SNe Ia data
yields ${\tilde{\Omega}}_{m}= 0.28\pm 0.01$ ($1\sigma$) where
$\tilde{{\Omega}}_{m}$ is the observed matter density parameter. In
particular, this implies that the  model has no dark energy but the
part of the matter that is effectively clustering is in good
agreement with the latest determinations from large scale structure.
The growth of perturbation and the formation of galaxy clusters in
such scenarios are also investigated. Despite the fact that
both scenarios may share the same Hubble expansion, we find that
matter creation cosmologies predict stronger small scale dynamics
which implies a faster growth rate of perturbations with respect to
the usual $\Lambda$CDM cosmology.  Such results point to the
possibility of a crucial observational test confronting CCDM with
$\Lambda$CDM scenarios trough a more detailed analysis involving
CMB, weak lensing, as well as the large scale structure.

\end{abstract}
\pacs{98.80.-k, 95.35.+d, 95.36.+x}
\keywords{Cosmology; accelerating Universe; dark matter; dark
energy}
\maketitle

\section{INTRODUCTION}
The analysis of high quality cosmological data (e.g. supernovae type
Ia, CMB, galaxy clustering, etc) have converged towards a cosmic
expansion history that involves a spatially flat geometry and some
sort of dark energy in order to explain the recent accelerating
expansion of the universe
\cite{Spergel07,essence,komatsu08,Teg04,Eis05,Kowal08,Hic09}. The
simplest dark energy candidate corresponds to a cosmological
constant, $\Lambda$ (see \cite{Weinberg89,Peebles03,Pad03} for
reviews). In the standard concordance  $\Lambda$CDM model, the
overall cosmic fluid contains baryons, cold dark matter plus a
vacuum energy that fits accurately the current observational data
and thus it provides an excellent scenario to describe the observed
universe.

On the other hand, the concordance model suffers from, among others
\cite{Peri08}, two fundamental problems: (a) {\it The fine tuning
problem} i.e., the fact that the observed value of the vacuum energy
density ($\rho_{\Lambda}=\Lambda c^{2}/8\pi G\simeq
10^{-47}\,GeV^4$) is more than 120 orders of magnitude below the
natural value estimated using quantum field theory
\cite{Weinberg89}, and (b) {\it the coincidence problem}
\cite{coincidence} i.e., the fact that the matter energy density and
the vacuum energy density are of the same order just prior to the
present epoch. Such problems have inspired many authors to propose
alternative candidates to dark energy such as $\Lambda(t)$
cosmologies, quintessence, $k-$essence, vector fields, phantom,
tachyons, Chaplygin gas and the list goes on (see
\cite{Reviews,Ratra88,Oze87,Free187,Wetterich:1994bg,Caldwell98,Brax:1999gp,
KAM,fein02,Caldwell,chime04,Brookfield:2005td,Bauer05,Grande06,
Boehmer:2007qa,Bas09b} and references therein).

Nowadays, the nature of the dark energy is considered one of the
most fundamental and difficult problems in the interface uniting
Astronomy, Particle Physics and Cosmology. However, there are other
possibilities. For instance, one may consider non-standard gravity
theories where the present accelerating stage of the universe is
driven only by cold dark matter, that is, with no appealing to the
existence of dark energy. Such a reduction of the so-called dark
sector is naturally obtained in the so-called $f(R)$ gravity
theories \cite{FR} (see, however, \cite{Nat10}).  Even in the
framework of the standard general relativity theory, is also
possible to reduce the dark sector by considering  the presence of
inhomogeneities \cite{Inhom}, quartessence models \cite{Quart}, as
well as the gravitationally induced particle creation mechanism
\cite{Parker,BirrellD,Prigogine,LCW,LG92,ZP2,LGA96,LA99,Susmann94,ZSBP01,SLC02,freaza02,Makler07,LSS08,SSL09,LJO09}.
In what follows we focus our attention to the last approach by
considering the  gravitational creation of cold dark matter in the
expanding Universe.

The basic microscopic description for gravitational particle production in
an expanding universe has been investigated in the literature by
many authors starting with Schr\"odinger \cite{SCHO39}.  In the late 1960s, independently of
Schr\"odinger's work, this issue was investigated by L. Parker and
others \cite{Parker,BirrellD} by considering  the Bogoliubov mode-mixing technique in the
context of quantum field theory in curved spacetime. The basic physical effect is that a classical
non-stationary background influences bosonic or fermionic quantum fields  in such a way that their masses become
time-dependent (see, e.g., \cite{Parker1} for more recent works). For a real scalar field in a flat Friedmann-Lema{\^i}tre-Robertson-Walker (FLRW) geometry described
in conformal coordinates, for example, the key result is that the
effective mass reads \cite{MW07},  $m^2_{eff}(\eta)\equiv m^2a^2-{a''/a}$, where m is
the ``Minkowskian" constant mass of the scalar field,
$a(\eta)$ is the scale factor, and the derivatives are computed with
respect to the conformal time. This time dependent  mass accounts
for the interaction between the scalar field and the geometry of the
Universe.  When the field is quantized, this leads to particle
creation, with the energy for newly created particles being supplied
by the classical, time-varying gravitational background.

Macroscopically, the first self-consistent formulation for matter
creation was put forward by Prigogine and coworkers \cite{Prigogine}
and somewhat clarified by Calv\~{a}o, Lima and Waga \cite{LCW} (see
also \cite{LG92}). It was shown that matter creation, at the
expenses of the gravitational field is also macroscopically
described by a negative pressure, and, potentially, can accelerate
the Universe. Within this framework, various interesting features of
cosmologies with particle creation have been discussed in
\cite{ZP2,LGA96,LA99,Susmann94,ZSBP01,SLC02,freaza02} (see also
\cite{Makler07} for recent studies on this subject). More recently,
the corresponding effects on the global dynamics of the particle
creation regime has been investigated extensively by a number of
authors (see \cite{LSS08,SSL09,LJO09} and references therein).  In
particular, it was also found that a subclass of such models
depends only of one free parameter and that the evolution of the
scale factor is exactly the same of $\Lambda$CDM models
\cite{LJO09}. Naturally, in order to know if creation of cold
dark matter (CCDM) models provide a realistic description of the
observed Universe, its viability need to be tested by discussing all
the constraints imposed from current observations both for background
and perturbative levels (structure formation).

In this context, the basic aim of the present work is twofold.
First, we place constraints on the main parameters of CCDM
cosmologies by performing a joint likelihood analysis involving  the
shift parameter of the Cosmic Microwave Background (CMB)
\cite{komatsu08} and the observed Baryonic Acoustic Oscillations
(BAO) \cite{Eis05}, and the latest SN Ia data (Constitution)
\cite{Hic09}. Secondly, for a wide class of matter creation
cosmologies, we also develop the linear approach for the density
perturbations. Analytical solutions for the differential equation
governing the evolution of the growth factor and some properties of
the large scale structure (cluster formation) are also discussed and
compared to the predictions of the $\Lambda$CDM model.

The paper is planned as follows. The basic theoretical elements of
the problem are presented in section 2, where we introduce  the
basic FLRW cosmological equations for CCDM cosmologies. In section 3
and 4 we present the specific CCDM  scenarios and derive the
constraints on their parameters based on a statistical joint analysis
involving the shift parameter of the Cosmic Microwave Background
(CMB)\,\cite{komatsu08}, the Baryonic Acoustic Oscillations (BAO)
\cite{Eis05}, and  the latest SN Ia data (Constitution)
\cite{Hic09}. In section 5 we study the evolution of linear
perturbations, while in section 6 we present the corresponding
theoretical predictions regarding the formation of the galaxy
clusters with basis on the Press-Schecther formalism. Finally, in
section 7 we draw our conclusions. Throughout the paper we consider
$H_{0}=70.5$ km/sec/Mpc as given by the WMAP 5-years data
\cite{komatsu08}.

\section{Creation Cold Dark Matter (CCDM) Cosmologies: Basic Equations}
The nontrivial cosmological equations for the mixture of radiation,
baryons and cold dark matter (with creation of dark matter
particles), and the energy conservation laws for each component have
been investigated thoroughly by
\cite{ZP2,LG92,LGA96,LSS08,SSL09,LJO09}. In this framework, for a
spatially flat FLRW geometry the basic equations which govern the
global dynamics of the universe in the matter dominated era
($\rho_{rad}\equiv 0$) are given by \cite{LSS08,SSL09,LJO09}

\begin{equation}
\label{fried} H^{2}= \frac{8\pi G}{3}(\rho_{bar}  + \rho_{dm}) =
\frac{8\pi G}{3}\rho_{m},
\end{equation}

\begin{equation}
\label{fried1} \frac{\ddot{a}}{a}= - \frac{4\pi G}{3}(\rho_{m}  +
3p_{c}),
\end{equation}

\begin{equation}\label{energy}
      \dot{\rho}_{bar}+ 3 H \rho_{bar}= 0,
\,\,\,\,\,\, \, \,\,\,\,\, \dot{\rho}_{dm} + 3 H \rho_{dm}=\rho_{dm}
\Gamma,
\end{equation}
which implies that
\begin{equation}
\label{ConsDM} \dot{\rho}_{m}+ 3 H \rho_{m}=\rho_{dm} \Gamma,
\,\,\,{\rm where}\;\; \rho_{m}=\rho_{dm}+\rho_{bar}.
\end{equation}
In the above set of differential equations, an over-dot denotes
derivative with respect to time, $\rho_{bar}$ and $\rho_{dm}$, are
the baryonic and dark matter energy densities, $H ={\dot {a}}/a$ is
the Hubble parameter, whereas $p_{c}$, corresponds to the creation
pressure. The quantity $\Gamma$ is the so called creation rate of
the cold dark matter and it has units of $(time)^{-1}$. The creation
pressure is defined in terms of the creation rate and other physical
quantities. For adiabatic creation of cold dark  matter, it has been
found \cite{LCW,LG92} that  the creation pressure is given by (see
also Prigogine et al. \cite{Prigogine})
\begin{equation}
\label{CP}
    p_{c} = -\frac{\rho_{dm} \Gamma}{3H}.
\end{equation}
Using Eqs. (\ref{fried}) and (\ref{ConsDM}) we can obtain the
following useful formula: \be \label{fried2}
\dot{H}+\frac{3}{2}H^{2}=\frac{4\pi
G\rho_{dm}}{3}\;\frac{\Gamma}{H}. \ee Expressions
(\ref{ConsDM})-(\ref{fried2}) show how the matter creation rate,
$\Gamma$, modifies the basic quantities of Einstein-de Sitter
cosmology. Note that if the creation rate is negligible, $\Gamma
<<H$, then, the creation pressure is  negligible, $\rho_m \propto
a^{-3}$,  and the solution of the above equation reduces to that of
the Einstein de-Sitter model, namely, $H(t)=2/3t$.

\section{TWO SPECIFIC CCDM Models}

Although the precise functional form of $\Gamma(t)$ is still
missing,  a number of different phenomenological parameterizations
have been proposed in the literature treating the time-dependent
$\Gamma(t)$ function (\cite{LSS08,SSL09,LJO09} and references
therein). In what follows, we consider two different flat matter
creation models namely Lima, Silva \& Santos (hereafter LSS
\cite{LSS08}) and Lima, Jesus \& Oliveira (hereafter LJO
\cite{LJO09}) respectively. With the aid of the current
observational data, we will also put stringent constraints on their
free parameters

\subsection{The LSS Model}
In the LSS scenario,  the functional form of $\Gamma$ is
phenomenologically parameterized  by the following linear expansion
in the Hubble parameter (see LSS for more details) \be \label{PS1}
\Gamma = 3\gamma H_{0} + 3\beta H. \ee It is worth noticing that in
its originally formulation by LSS, both the baryonic and radiative
contributions were neglected ($\rho_{bar}=\rho_{rad}=0$). Such
incompleteness  was cured in a subsequent paper by Steigman et al.
\cite{SSL09}. The advantage of the LSS model is that it is
analytically described and due to this feature  we will consider it
as an interesting and preparatory toy model for more physical CCDM
scenarios.

The mass density in LSS reduces to the dark matter density,
$\rho_{m}=\rho_{dm}$ (see Eq. \ref{ConsDM}). Without going into the
details of that model, let us present its main  features. Based
on the latter considerations the basic Friedmann equation becomes
(see Eq. \ref{fried2}) \be \label{fried3}
\dot{H}+\frac{3}{2}H^{2}(1-\beta-\frac{\gamma H_{0}}{H})=0. \ee
Performing the integration of Eq.(\ref{fried3}) we derive the
following Hubble function: \be
H(t)=H_0\left(\frac{\gamma}{1-\beta}\right)\frac{e^{\frac{3\gamma
H_0}{2}t}}{(e^{\frac{3\gamma H_0}{2}t}-1)},
\label{frie55} \ee and by integrating the latter we obtain the
evolution of the scale factor \be a(t)= {1 \over (1 + z)} =
\left[\left({1 - \gamma - \beta \over \gamma}\right) ({\rm
e}^{3\gamma H_{0}t/2} - 1)\right]^{{2 \over 3(1 - \beta)}}
\label{frie5}. \ee Evaluating Eq.(\ref{frie5}) at the present time
($a\equiv1$ or $z=0$), we can define the present age of the Universe
\be t_{0} = {2 \over 3\gamma H_{0}}{\rm ln}\left({1 - \beta \over 1
- \gamma - \beta}\right). \ee

Now utilizing equations (\ref{frie55}) and (\ref{frie5}), we find
after some algebra, that the normalized Hubble flow takes the
following form \cite{SSL09} \be
E(a)=\frac{H(a)}{H_{0}}=\frac{\gamma}{1-\beta}+
\frac{1-\gamma-\beta}{1-\beta}a^{-3(1-\beta)/2}. \ee We now study
the conditions under which an inflection point exists in our past,
implying an acceleration phase of the scale factor. This crucial
period in the cosmic history corresponds to $\ddot{a}(t_{I})=0$ and
$t_{I}<t_{0}$. Differentiating twice Eq. (\ref{frie5}), we simply
have: \be a_{I}=\frac{1}{1 + z_{I}}= \left[{2\gamma \over (1 -
3\beta)(1 - \gamma - \beta)}\right]^{{-2 \over 3(1- \beta)}}. \ee
Note that if $\gamma = 0$, there is no transition from an early
decelerating to a late time accelerating Universe \cite{SSL09}. In
this case, one can also prove that the expansion of the Universe
always decelerates for $0 \leq \beta < 1/3$ and always accelerates
for $1/3 < \beta \leq 1$. The latter condition implies that the
parameter $\gamma$ has to obey the following restriction $0 < \gamma
\leq 1$. When $\gamma=0$, negative values of $\beta$ means
destruction of particles, and, for completeness, such a possibility
will also be considered in the discussion of some background tests
(see section 4).

\subsection{The LJO Model}

Now, let us consider an alternative CCDM scenario that includes the
baryonic and  radiative components, and was also proposed with the
intention to solve the cluster problem in such a  framework
\cite{LJO09}. In this context, the particle creation rate reads
\begin{equation}
\label{PS2}
\Gamma=3\tilde{\Omega}_{\Lambda}\left(\frac{\rho_{co}}{\rho_{dm}}\right)H,
\end{equation}
where $\tilde{\Omega}_{\Lambda}$ (called $\alpha$ in the
\cite{LJO09}) is a constant, $\rho_{co}=3H^{2}_{0}/8\pi G$ is the
present day value of the critical density, and the factor 3 has been
maintained for mathematical convenience. We stress that such CCDM
scenario mimics the global dynamics of the traditional $\Lambda$
cosmology \cite{LJO09}. Indeed, the combination of equations
(\ref{fried}), (\ref{fried2}),  and (\ref{PS2}) leads to the
following formula: \be \label{fried4}
\dot{H}+\frac{3}{2}H^{2}=\frac{3}{2}\;\tilde{\Omega}_{\Lambda}
H^{2}_{0}, \ee a solution of which is given by
\begin{equation}
H(t)=\sqrt{\tilde{\Omega}_{\Lambda}}\;H_{0} \;{\rm
  coth}\left(\frac{3 H_{0}\sqrt{\tilde{\Omega}_{\Lambda}}}{2}\;t\right).
\label{frie556}
\end{equation}
Using now the definition of the Hubble parameter $H\equiv {\dot
a}/a$, the scale factor of the universe $a(t)$, normalized to unity
at the present epoch, evolves with time as:
\begin{eqnarray}
\label{all}
a(t)=\left(\frac{\tilde{\Omega}_{m}}{\tilde{\Omega}_{\Lambda}}\right)^{1/3}
\sinh^{\frac{2}{3}}\left(\frac{3H_{0}\sqrt{\tilde{\Omega}_{\Lambda}}
}{2}\; t\right).
\end{eqnarray}
where $\tilde{\Omega}_{m}=1-\tilde{\Omega}_{\Lambda}\equiv 1 -
\alpha$, can be viewed as the effective matter density parameter at
the present time. For an arbitrary redshift,
$\tilde{\Omega}_{m}(z)$, measures the mass density that is
effectively  clustering \cite{LJO09}. The cosmic time is related
with the scale factor as
\begin{equation}\label{time}
t(a)=\frac{2}{3\sqrt{\tilde{\Omega}_{\Lambda}}H_{0}  } {\rm
sinh^{-1}} \left(\sqrt{ \frac{\tilde{\Omega}_{\Lambda}}
{\tilde{\Omega}_{m}}} \;a^{3/2} \right).
\end{equation}
Combining the above equations we can define the normalized Hubble
expansion as a function of the scale factor:
\begin{eqnarray}
\label{hub1} E(a)=\frac{H(a)}{H_{0}}= \left(\tilde{\Omega}_{m}a^{-3}
+ \tilde{\Omega}_{\Lambda} \right)^{1/2}.
\end{eqnarray}
Obviously, the LJO model contains only one free parameter. The
inflection point [namely, the point where the Hubble expansion
changes from the decelerating to the
  accelerating regime, $\ddot{a}(t_{I})=0$] takes place at:
\begin{eqnarray}
      \label{infle}
t_{I}=\frac{2}{3\sqrt{\tilde{\Omega}_{\Lambda}}H_{0}} {\rm
sinh^{-1}} \left(\sqrt{ \frac{1} {2}} \right) \;,\;\;
a_{I}=\left[\frac{\tilde{\Omega}_{m}}{2\tilde{\Omega}_{\Lambda}}\right]^{1/3}.
\end{eqnarray}
In what follows we constrain the free parameters of such models
coming from the background tests. For the sake of clarity, the study
of the perturbed models and the associated constraints coming from
galaxy cluster formation will be postpone to sections 5 and 6.

\section{Background Tests: BAO, CMB Shift Parameters and SNe Ia data}
In this section we briefly present the statistical analysis used in
order to constrain the matter creation models presented in the
previous section. First of all, we consider the Baryonic Acoustic
Oscillations (BAO). This kind of estimator is  produced by pressure
(acoustic) waves in the photon-baryon plasma in the early universe,
generated by dark matter overdensities. Evidence of this excess was
recently found in the clustering properties of the luminous SDSS
red-galaxies \cite{Eis05}  and it provides a ``standard ruler'' with
which we can constrain dark energy models. In particular, we use the
following estimator: $ A({\bf
p})=\frac{\sqrt{\Omega_{m}}}{[z^{2}_{s}E(a_{s})]^{1/3}}
\left[\int_{a_{s}}^{1} \frac{da}{a^{2}E(a)} \right]^{2/3}$, measured
from the SDSS data to be $A=0.469\pm 0.017$, where $z_{s}=0.35$ [or
$a_{s}=(1+z_{s})^{-1}\simeq 0.75$]. Therefore, the corresponding
$\chi^{2}_{\rm BAO}$ function is simply written
\begin{equation}
\chi^{2}_{\rm BAO}({\bf p})=\frac{[A({\bf p})-0.469]^{2}}{0.017^{2}}
\end{equation}
where ${\bf p}$ is a vector containing the cosmological parameters
that we want to fit.

On the other hand, a very accurate and deep geometrical probe of
dark energy is the angular scale of the sound horizon at the last
scattering surface as encoded in the location $l_1^{TT}$ of the
first peak of the Cosmic Microwave Background (CMB) temperature
perturbation spectrum. This probe is described by the  CMB shift
parameter \cite{Bond:1997wr,Nesseris:2006er} and is defined as $
R=\sqrt{\Omega_{m}}\int_{a_{ls}}^1 \frac{da}{a^2 E(a)}$. The shift
parameter measured from the WMAP 5-years data \cite{komatsu08} is
$R=1.71\pm 0.019$ at $z_{ls}=1090$ [or $a_{ls}=(1+z_{ls})^{-1}\simeq
9.17\times 10^{-4}$]. In this case, the $\chi^{2}_{\rm cmb}$
function is given
\begin{equation}
\chi^{2}_{\rm cmb}({\bf p})=\frac{[R({\bf p})-1.71]^{2}}{0.019^{2}}
\end{equation}
Note that the measured CMB shift parameter is somewhat model
dependent but mostly to models which are not included in our
analysis. For example, in the case where massive neutrinos are
included or when there is a strongly varying equation of state
parameter. The robustness of the shift parameter was tested and
discussed in \cite{Elgaroy07}.

Finally, we use the 'Constitution set' of 397 type Ia supernovae
of Hicken et al. \cite{Hic09}. In order to avoid possible problems
related with the local bulk flow, we use a subsample of the overall
sample in which we select those SNIa with $z>0.023$. This subsample
contains 351 entries. The corresponding $\chi^{2}_{\rm SNIa}$
function becomes:
\begin{equation}
\label{chi22} \chi^{2}_{\rm SNIa}({\bf p})=\sum_{i=1}^{351} \left[
\frac{ {\cal \mu}^{\rm th} (a_{i},{\bf p})-{\cal \mu}^{\rm
obs}(a_{i}) } {\sigma_{i}} \right]^{2} \;\;.
\end{equation}
where $a_{i}=(1+z_{i})^{-1}$ is the observed scale factor of the
Universe, $z_{i}$ is the observed redshift, ${\cal \mu}$ is the
distance modulus ${\cal \mu}=m-M=5{\rm log}d_{L}+25$ and
$d_{L}(a,{\bf p})$ is the luminosity distance $ d_{L}(a,{\bf
p})=\frac{c}{a} \int_{a}^{1} \frac{{\rm d}y}{y^{2}H(y)} $ where $c$
is the speed of light. We can combine the above probes by using a
joint likelihood analysis: ${\cal L}_{tot}({\bf p})= {\cal L}_{\rm
BAO} \times {\cal L}_{\rm cmb}\times {\cal L}_{\rm SNIa}$ or
$\chi^{2}_{tot}({\bf p})=\chi^{2}_{\rm BAO}+\chi^{2}_{\rm
cmb}+\chi^{2}_{\rm SNIa}$ in order to put even further constraints
on the parameter space used\footnote{Likelihoods are normalized to
their maximum values. We will report $1\sigma$ uncertainties of the
fitted parameters. The overall number of data points used is
$N_{tot}=353$ and the degrees of freedom: {\em
  dof}$= N_{tot}-n_{\rm fit}$, with $n_{\rm fit}$ the number of fitted
parameters, which varies for the different models.}.

Performing now our statistical analysis we attempt to put constrains
on the free parameters. In particular, we sample $\beta \in
[-0.2,0.2]$ and $\gamma \in [0.1,1]$ in steps of 0.001. Thus, the
overall likelihood function peaks at $\beta=-0.15 \pm 0.007$ and
$\gamma=0.86 \pm 0.01$ with $\chi_{tot}^{2}(\beta,\gamma)=432.5$
({\em dof}=$351$). Using the latter cosmological parameters the
corresponding current age of the universe is found to be $t_{0}\sim
14.8$ Gyr while the inflection point is located at $a_{I}\simeq
0.44$. This corresponds to a redshift $z_I\simeq 1.26$, which is
substantially higher than in the case of the concordance model. If
we now impose $\beta=0$, the SNIa likelihood analysis provides a
best fit value of $\gamma=0.66 \pm 0.02$, in agreement with
\cite{SSL09}. Note that \cite{SSL09} used only the SNIa data. We
find that although our SNIa likelihood analysis provides a similar
$\gamma$ value to that found in \cite{SSL09}, our joint likelihood
funtion peaks at $\gamma=0.41 \pm 0.01$ but with a poor quality fit:
the reduced $\chi_{tot}^{2}$ is $\simeq 3$.

This simply means that the functional form
$E(a)=\gamma+(1-\gamma)a^{-3/2}$ is unable to fit the observational
data simultaneously at low and high redshifts. We confirm this point
by using the CMB shift parameter only, finding that the
corresponding likelihood function peaks at $\gamma \simeq 0.20$.
This value is $\sim 3.4$ times larger than that provided by the
SNIa+BAO solution $\gamma \simeq 0.68$ (SNIa only points $\gamma
\simeq 0.66$). This fact alone suggests that for the class of models
with a constant creation rate $\Gamma=3\gamma H_{0}$, are unable to
provide a quality fit of the basic cosmological data in all the
relevant redshift ranges.

Now, let us constrain the LJO model with the observational data from
background tests. From our joint analysis involving BAO + CMB Shift
Parameter + SNe Ia data,  we find that the best fit value (within
$1\sigma$ uncertainty) is $\tilde{\Omega}_{m}=0.28\pm 0.01$ with
$\chi_{tot}^{2}(\Omega_{m})\simeq 431.5$ ({\em dof}=$352$). This
result is in good agreement with recent studies
\cite{Spergel07,essence, komatsu08, Kowal08, Hic09} of the
concordance $\Lambda$ cosmology. Therefore, the current age of the
universe is $t_{0}\simeq 13.9$Gyr while the inflection point is
located at $a_{I}\simeq 0.58$ (hence at redshift $z_I\simeq 0.72$).

Summing up, although solving the acceleration and age problems,
the LSS type scenarios is endowed with severe difficulties even for
the set of background tests analysed here.  On the other hand, the
LJO scenario provides good fits for SNe + BAO + CMB shift parameter.
A more detailed analysis on this issue (background tests) it will be
published in a forthcoming paper.

\section{Matter Density Perturbations in CCDM Cosmologies}

Let us now derive the basic equation that governs the evolution of
the mass density contrast, modeled here as an ideal
fluid in a `quasi'-Newtonian or neo-Newtonian framework \cite{LZB97}.
In this approach, background equations are formulated in a way that the isotropic
pressure becomes dynamically relevant even for FLRW cosmologies where
all the physical properties may depend only of time.  This
allows us to describe an accelerated expansion of the Universe powered by an
effective negative pressure in a Newtonian framework. Naturally, even considering that the
neo-Newtonian approach reproduces the GR background dynamics
exactly, small differences may occur at the perturbative level.
However, as discussed by Reis \cite{Reis03}, the GR
first-order perturbation dynamics and its neo-Newtonian counterpart
coincide exactly for a vanishing sound speed. Indeed, for
constant equations of state it has been demonstrated
that the correct large-scale behavior in the synchronous gauge are also
reproduced. This should also be expected since the observational data
correspond to modes that are well inside the
Hubble radius, for which the use of a neo-Newtonian approach seems to be
adequate (for recent papers in this subject see \cite{FZ} and
references therein)

\subsection{Basic Formalism}

In what follows we will adopt a nonrelativistic description with basis in   
an extended continuity equation together the Euler and
Poisson equations. In virtue of  the particle creation process they
take the form:

\be \label{cons} \left(\frac{\partial \rho}{\partial t}
\right)_{r}+{\bf \nabla}_{r} \cdot (\rho {\bf u})= \rho \Gamma, \ee
\be \label{euler} \left(\frac{\partial {\bf u}}{\partial t}
\right)_{r}+({\bf u} \cdot {\bf \nabla}_{r}) {\bf u}=-{\bf \nabla}
\Phi, \ee and \be \label{poiss} \nabla^{2}_{r}\Phi=  4\pi G \rho
-\Lambda_{eff}, \ee where $({\bf r},t)$ are the proper coordinates,
${\bf u}$ is the velocity of a fluid element of volume, $\rho$ is
the mass density and $\Phi$ is the gravitational potential. Since we
are working within the context of cosmological models with matter
creation we have modified the continuity equation (\ref{cons}) by
including the "standard"
source term. Utilizing Eq. (\ref{fried1})
and Eq. (\ref{CP}), the quantity $\Lambda_{eff}$ is defined as \be
\label{poiss1}
\Lambda_{eff}= \Lambda -12\pi G p_{c}= \Lambda +4\pi
G \rho_{dm} \frac{\Gamma}{H},
\ee
where for completeness and  future
comparison to the present cosmic concordance model  ($\Lambda$CDM),
we have also included the cosmological $\Lambda$-term.

Now, by  changing from proper $({\bf r},t)$ to comoving $({\bf
x},t)$ variables, the  fluid velocity becomes \cite{Pee93} \be
\label{vel} {\bf u}=\dot{a} {\bf x}+a {\bf \dot {x}}=\dot{a} {\bf
x}+{\bf v}({\bf x},t), \ee while the corresponding differential
operators take the following forms \be \label{oper1} {\bf
\nabla}_{x} \equiv {\bf \nabla}=a {\bf \nabla}_{r}, \ee and \be
\label{oper2} \left(\frac{\partial }{\partial t} \right)_{x} \equiv
\frac{\partial }{\partial t}= \left(\frac{\partial }{\partial t}
\right)_{r}+H{\bf x}\cdot \bf{\nabla}, \ee where ${\bf x}={\bf r}/a$
and ${\bf v}({\bf x},t)$ is the peculiar velocity with respect to
the general expansion. Note that the mass density is written as \be
\label{mass1} \rho=\rho_{m}(t)[1+\delta({\bf x},t)]. \ee In this
context, using eqs.(\ref{ConsDM}), (\ref{oper1}) and (\ref{oper2})
and neglecting second order terms ($\delta \ll 1$ and $v \ll u$) it
is routine to rewrite eqs.(\ref{cons}), (\ref{euler}) and (\ref
{poiss}) \be \ddot{a} {\bf x}+\frac{\partial {\bf v}}{\partial t}+H
{\bf v}= -\frac{{\bf \nabla}  \Phi}{a}, \ee \be \label{avell} {\bf
{\nabla} \cdot v}=-a\left(\frac{\partial \delta}{\partial
t}+\frac{\psi \delta}{\rho_{m}} \right), \ee \be
\frac{1}{a^{2}}\nabla^{2} \Phi=4\pi G
\rho_{m}(1+\delta)-\Lambda_{eff}. \ee Following the notations of
\cite{Pee93}, we write the gravitational potential as \be
\Phi=\phi({\bf x},t)+\frac{2}{3}\pi G \rho_{m} a^{2}
x^{2}-\frac{1}{6}\Lambda_{eff} a^{2}x^{2}. \ee Thus utilizing
Eqs.(\ref{fried}), (\ref{CP}) and (\ref{poiss1}) we derive after
some algebra that \be \label{vell1} \frac{\partial {\bf v}}{\partial
t}+H {\bf v}= -\frac{{\bf \nabla}  \phi}{a}, \ee and \be
\label{poiss2} \nabla^{2} \phi=4\pi G a^{2} \rho_{m} \delta. \ee
Finally, by taking the divergence of Eq. (\ref{vell1}) and using
Eqs. (\ref{avell}) and (\ref{poiss2}), we obtain the time evolution
equation for the growth factor $D(t)\equiv\delta/A({\bf x})$ [where
$A({\bf x})$ is an arbitrary function]
\begin{equation}
\label{eq:11} \ddot{D}+(2H+Q)\dot{D}-\left(4\pi G \rho_{m}
-2HQ-\dot{Q} \right)D=0,
\end{equation}
where \be \label{eq:12} Q(t)=\frac{\rho_{dm} \Gamma}{\rho_{m}}. \ee
Now, it is clear how the matter creation term affects the growth
factor via the function $Q(t)$.

It should be noticed that the background creation pressure affects
both the global
dynamics as well as the fluctuations via $Q(t)$.
In the above results, it was also implicitly assumed that
the produced particles have negligible velocities
as measured by the observers in the comoving framework which implies
that any additional internal pressure term is negligible. In fact this
is a reasonable assumption since we are working with a
non-relativistic phase of universe expansion (matter era). The second
assumption is that the matter production is strictly homogeneous
which implies that possible contributions of $Q(t)$ at the
perturbative level is practically zero.
Thus, within the context of the latter assumptions, the advantage of
employing a neo-Newtonian approach
is a gain in simplicity and transparency of all computations.

In order to solve the
above differential equation we need to know the functional form of
the quantity $Q(t)$ (or $\Gamma$). In the case of a 
negligible matter creation rate, $\Gamma<<H$ [$Q(t)
\sim 0$], the above equation reduces to the usual time evolution
equation for the mass density contrast \cite{Pee93} for  which  the
growth factor is $D_{EdS}(a)=a$. As one may check,  by solving Eq.
(\ref{eq:11}) for the $\Lambda$ cosmology [$Q(t)=0$], we also derive
the well known perturbation growth factor (see \cite{Pee93}):
\be\label{eq24} D_{\Lambda}(a)=\frac{5\Omega_{m}
  E(a)}{2}\int^{a}_{0}
\frac{dx}{x^{3}E^{3}(x)}, \ee where $\Omega_{m}=1-\Omega_{\Lambda}$
is the matter density parameter and $E(a)=H(a)/H_{0}$ is the
normalized Hubble function
\begin{eqnarray}
\label{hubl} E(a)=\left(\Omega_{\Lambda}+\Omega_{m}a^{-3}
\right)^{1/2}.
\end{eqnarray}
It is also interesting to mention here that one can explicitly
derive Eq. (\ref{eq:11}) in the framework of cosmological models
with a time varying vacuum energy density \cite{Arc94}. Note, that
for simplicity throughout the rest of the paper we will use
geometrical units ($c=8\pi G \equiv 1$). In the analytical treatment
below,  unless explicitly stated, we consider $\Lambda=0$.

\subsection{Growth of Fluctuations in CCDM Models: Analytical Solutions}

Let us now discuss thoroughly the time evolution equation of the
mass density contrast in the linear regime in order to test
the implications of CCDM models on the structure formation process.
In order to illustrate our results with a basic application, we also
consider later the formation of galaxy clusters through the
so-called Press-Schetcher formalism.

\subsubsection{Fluctuations  in LSS Model}
Using the time evolution equation for the mass density contrast as
given by (\ref{eq:11}), we now derive the growth factor of
fluctuations for LSS cosmology. To begin with,  we change variables
from $t$ to a new one defined by the following transformation \be
\label{tran1} y={\rm e}^{n t/2}-1 \;\;{\rm with} \;\; n=3\gamma
H_{0}. \ee In this context, using equations (\ref{fried}),
(\ref{ConsDM}), (\ref{PS1}), (\ref{frie55}), (\ref{frie5}),
(\ref{eq:12}) and (\ref{tran1}) we obtain
\begin{eqnarray}
\label{auxiliar1}
H(y)=\frac{n (y+1)}{\mu y},\nonumber\\
\rmm(y)=\frac{3n^{2} (y+1)^{2}}{\mu^{2} y^{2}},\nonumber\\
Q(y)=\frac{n[3y-(\mu-3)]}{\mu y},\nonumber\\
\dot{Q}(y)=\frac{n^{2} (\mu-3)(y+1)}{2\mu y^{2}},
\end{eqnarray}
where $\mu=3(1-\beta)$. We can now rewrite Eq. (\ref{eq:11}) as:
\be
\label{eq:22}
\mu^{2} (y+1)y^{2}D^{''}+\mu y f(y)D^{'}+2g(y)D=0,
\ee
where prime denotes derivatives with respect to $y$ and
\begin{eqnarray}\label{fgpsi1}
f(y)&=&(10+\mu)(y+1)-3\mu \nonumber\\
g(y)&=&9(y+1)-7\mu+\mu^{2}\;.
\end{eqnarray}
Factors of $n$ drop at the end of the calculation. Notice, that this
variable is related with the scale factor as \be \label{tra2}
y=\frac{3\gamma}{\mu-3\gamma}\,a^{\mu/2}. \ee
We find that eq.(\ref{eq:22}) has a growing solution of the form
\begin{equation}
\label{eqff2} D(y)={\cal C} y^{(-10+3\mu+\sqrt{\Delta})/2\mu}
F\left(\nu_1,\nu_1+\frac{2\sqrt{7}}{\mu}, \nu_{2},-y \right),
\end{equation}
where
\begin{eqnarray}
\nu_{1}=\frac{3}{2}-\frac{\sqrt{7}}{\mu}+\frac{\sqrt{\Delta}}{2\mu}
\;\;\;
\nu_{2}=1+\frac{\sqrt{\Delta}}{\mu},\nonumber\\
\Delta=28-4\mu+\mu^{2}.
\end{eqnarray}
Notice that the quantity $F$ is the hypergeometric function and
${\cal C}$ is a constant. Inserting eq.(\ref{tra2}) into
eq.(\ref{eqff2}), we obtain the growth factor $D(a)$ as a function
of the scale factor:
\begin{equation}
\label{eqff1} D(a)=C_{1} a^{(-10+3\mu+\sqrt{\Delta})/4}
F\left(\nu_1,\nu_1+\frac{2\sqrt{7}}{\mu}, \nu_{2},-\frac{3\gamma
a^{\mu/2}}{\mu-3\gamma} \right),
\end{equation}
where
\begin{equation}
C_{1}={\cal C} \, \left(\frac{3\gamma}{\mu-3\gamma}
\right)^{(-10+3\mu+\sqrt{\Delta})/2\mu}.
\end{equation}

\subsubsection{Fluctuations in LJO Models}

Let us now proceed further to solve  analytically the differential
Eq. (\ref{eq:11}) in order to investigate the matter fluctuation
field of the LJO model in the linear regime. To do so, we change
variables from $t$ to $y$ according to the transformation
\be\label{tran11} y={\rm
  coth}\left(\frac{3 H_{0}\sqrt{\tilde{\Omega}_{\Lambda}}}{2}\;t\right).
\ee
Using  (\ref{tran11}) we find, after some calculations, that
equation (\ref{eq:11}) takes on the form
\be \label{eq:222} 3y^{2}(y^{2}-1)^{2}D^{''}+2yf(y) D^{'}- 2g(y)D=0,
\ee
where
\begin{eqnarray}\label{fgpsi}
f(y)=(y^{2}-1)(y^{2}-3),\, \, g(y)=(y^{4}-7y^{2}+3).
\end{eqnarray}
Note that factors of $H_{0}$ drop at the end of the calculation.

In deriving Eq.(\ref{eq:222}) we have substituted the various terms
in (\ref{eq:11}) as a function of the new variable [see Eq.
(\ref{tran11})]. For instance, from equations  (\ref{fried}),
(\ref{frie556}) and (\ref{eq:12}), we have
\begin{eqnarray}\label{auxiliar}
H(y)= \sqrt{\tilde{\Omega}_{\Lambda}}H_{0}\,y,\nonumber\\
\rmm(y)= 3\,\tilde{\Omega}_{\Lambda}H^{2}_{0}\,y^{2},\nonumber\\
Q(y)=\frac{3\sqrt{\tilde{\Omega}_{\Lambda}}H_{0}}{y},\nonumber\\
\dot{Q}(y)=\frac{9\sqrt{\tilde{\Omega}_{\Lambda}}H_{0}\;(y^{2}-1)}{2y^{2}}.
\end{eqnarray}

In this framework, the differential equation (\ref{eq:222}) can be
brought into the following expression \be \label{eq:223}
6x^{2}(x+1)^{2}\;\frac{d^{2}D}{dx^{2}}+xq(x)\;\frac{dD}{dx}-b(x)D=0,
\ee
\begin{eqnarray}\label{fgpsi1}
q(x)=(x+1)(5x-4) \;\; b(x)=(x^{2}-5x-3),
\end{eqnarray}
where
\begin{equation}
\label{pps11} x=y^{2}-1=\left( \frac{\tilde{\Omega}_{m}}
{\tilde{\Omega}_{\Lambda}}\right)\;a^{-3} \;,\;\;\;\;\;\;
x+1=y^{2}=\frac{E^{2}(a)}{\tilde{\Omega}_{\Lambda}},
\end{equation}
in which we have used equations (\ref{frie556}), (\ref{all}) and
(\ref{hub1}).

The solution of the main differential equation (\ref{eq:223}) is
\begin{equation}
\label{DRG} D(x)={\cal C}\frac{x^{(5-\sqrt{7})/6}}{x+1}
F\left[\kappa,\kappa+\frac{5}{6}, 2(\kappa+\frac{7}{6}),-x \right],
\end{equation}
where $\kappa=-(4+\sqrt{7})/6$ and ${\cal C}$ is a constant.
Inserting Eq. (\ref{pps11}) into Eq. (\ref{DRG}), we finally obtain
the growth factor $D(a)$ as a function of the scale factor:
\begin{equation}
\label{DRG1} D(a)=C_{1}\;\frac{a^{-(5-\sqrt{7})/2}}{E^{2}(a)}
F\left[\kappa,\kappa+\frac{5}{6},
2(\kappa+\frac{7}{6}),-\frac{\tilde{\Omega}_{m}}{\tilde{\Omega}_{\Lambda}}\,\,a^{-3}
\right],
\end{equation}
where
\begin{equation}
C_{1}={\cal C}
\,\tilde{\Omega}_{\Lambda}\left(\frac{\tilde{\Omega}_{m}}
{\tilde{\Omega}_{\Lambda}} \right)^{(5-\sqrt{7})/6}.
\end{equation}
We would like to stress here that the above solution of the growth
factor is valid when $\tilde{\Omega}_{\Lambda} \ne 0$ (or
$\tilde{\Omega}_{m} \ne 1$).

\subsection{Linear Growth: Analysis and Summary}

Let us now discuss the evolution of the growth factor as a function
of the redshift [$D(z)$, $z=a^{-1}-1$]  as predicted by the two CCDM
cosmologies (see Eqs. (\ref{eqff1}) and (\ref{DRG1})). For
comparison we also present the growth factor provided by the
standard  $\Lambda$CDM cosmology (see Eq. \ref{eq24}).

In figure 1 we display the basic results. First, we notice that the
growth factors are normalized to unity at the present time ($z=0$).
It is also evident that the LSS growth factor (dashed line) is much
greater with respect to the other 2 models, LJO (dot line) and
$\Lambda$ (solid line). In this framework, we find that for $z\le
0.8$ the LSS growth factor (dashed line hereafter LSS$_{1}$) starts
to decay, implying that cosmic structures cannot be formed via
gravitational instability. The same general behavior seems to hold
also for the LSS model with $\beta=0$ and $\gamma=0.66$ (dot-dashed
line-hereafter LSS$_{2}$).

\begin{figure}[t]
\mbox{\epsfxsize=10cm \epsffile{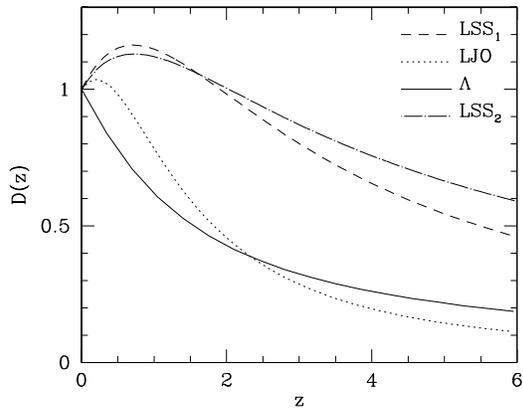}} \caption{ The evolution of
the growth factor. The lines correspond to LSS$_{1}$ [dashed,
$(\beta,\gamma)=(-0.15,0.86)$], LSS$_{2}$ [dot-dashed,
$(\beta,\gamma)=(0,0.66)$], LJO (dot) and $\Lambda$ cosmology
(solid). Note that the growth factors of all models were scaled to
unity at the present time.  Despite the fact that LJO and
$\Lambda$CDM models share the same global dynamics, wee see that
they trace differently  the evolution of the matter fluctuation
field (compare the dotted and solid lines).}
\end{figure}

Let us now compare the predictions of LJO and $\Lambda$CDM
growth factors. First we stress a remarkable result: despite the
fact that the two models share the same global dynamics (compare
Eqs. (\ref{hub1}) and (\ref{hubl})), they trace differently the
evolution of the matter fluctuation field. In particular, close to
the present epoch ($z<0.4$) the LJO growth factor (dot-dashed line)
reaches almost a plateau, which means that the matter fluctuations
are effectively frozen. Between $0.4 \le z < 1.6$, the growth factor
in the LJO model is greater than that of the concordance
$\Lambda$CDM cosmology. In particular, the corresponding deviation
$(1-D/D_{\Lambda})\%$ lies in the interval $[-35,-19]\%$. As an
example, prior to the inflection point ($z_{I}\sim 0.72$) we find
$\sim -28\%$.

It is worth noting that in the interval of $1.6 \le z \le 2.8$ the
LJO growth factor tends to the $\Lambda$ solution. Indeed, at the
epoch of $z\sim 1.62$ ($a\sim 0.38$), in which the most distant
cluster has been found \cite{Pap10}, the deviation
$(1-D/D_{\Lambda})\%$, of the growth factor $D(z)$ for the LJO
scenario with respect to the $\Lambda$ solution $D_{\Lambda}(z)$ is
$-5\%$. To this end, for $z>2.8$ we find that $D(z)<D_{\Lambda}(z)$
and the corresponding deviation $(1-D/D_{\Lambda})\%$, remains close
to $\sim 28\%$. Naturally, one may expected that the
differences among the growth factors will affect the predictions
related with the formation of the cosmic structures (see next
section). In particular, such results point to the possibility of a
crucial observational test confronting CCDM with $\Lambda$CDM
scenarios in a perturbative level.

\section{The formation and evolution of collapsed structures}
In this section we study the cluster formation processes in
CCDM cosmologies  by using the standard Press-Schecther formalism
\cite{press}. In such an approach, the behavior of the matter
perturbations is described by assuming that the density contrast
behaves like a Gaussian random field.  As it is widely known, the
cluster distribution basically traces scales that have not yet
undergone the non-linear phase of gravitationally clustering, thus
simplifying their connections to the initial conditions of cosmic
structure formation.

The basic aim here is to estimate the fractional rate of cluster
formation (see \cite{Peeb84,Rich92}). In particular, these studies
introduce a methodology which computes the rate at which mass joins
virialized halos (such as galaxy clusters) which grow from small
initial perturbations in the universe,  with matter
fluctuations, $\delta$, greater than a critical value $\delta_{c}$.

Now, by assuming that the density fluctuation field, smoothed at the
scale $R$ (corresponding to a mass scale of $M=4\pi\bar{\rho}
R^3/3$, with $\bar{\rho}$ the mean background mass density of the
Universe), is normally distributed with zero mean, then the
probability that the field will have a value $\delta$ at any given
point in space is: \be\label{eq:88} {\cal P}(\delta, z)
=\frac{1}{\sqrt{2\pi}\sigma(R,z)} {\rm
exp}\left[-\frac{\delta^{2}}{2\sigma^{2}(R,z)} \right], \ee where
the variance of the Gaussian field, $\sigma^2(R,z)$, is given by:
\be \sigma^2(R,z) = \frac{1}{2 \pi^2} \int_0^{\infty} k^2 P(k,z)
W^2(kR) dk, \ee with $P(k,z)$ the power spectrum of density
fluctuations which evolves according to $P(k,z)=P(k,0) D^2(z)$, with
$D(z)$ the growing mode of the density fluctuations evolution,
normalized such that $D(0)=1$.

Finally, $W(kR)$ is the top-hat smoothing kernel, given in
Fourier-space by: \be W(kR)=\frac{3}{(kR)^3} [\sin(kR)-kR \cos(kR)].
\ee
We can now estimate what fraction of the Universe, at some reference
redshift $z$ and above some mass threshold $M$, has collapsed to
form bound structures. To this end we need to integrate the
probability function given by Eq. (\ref{eq:88}) over all regions
that at some prior redshift had overdensities which by the reference
redshift have increased to above the critical value, $\delta_c(z)$,
which in an Einstein-deSitter universe is $\simeq 1.686$ and varies
slightly for different values of $\Omega_m$ \cite{LaceyCole,eke}.
Therefore, this fraction is given by
\cite{Rich92},\cite{Bas1},\cite{Bas2},\cite{Bas3}:

\be {\cal F}(M,z)=\int_{\delta_{c(z)}}^{\infty} {\cal P}(\delta, z)
d\delta, \ee and performing the above integration, parametrizing the
rms mass fluctuation amplitude at $R=8 \; h^{-1}$ Mpc which can be
expressed as a function of redshift as
$\sigma(M,z)=\sigma_{8}(z)=D(z)\sigma_{8}$, we obtain: \be
\label{eq:89} {\cal F}(z)=\frac{1}{2} \left[1-{\rm erf} \left(
\frac{\delta_{c}}{\sqrt{2} \sigma_{8}(z)} \right) \right]. \ee

\begin{figure}[ht]
\mbox{\epsfxsize=10cm \epsffile{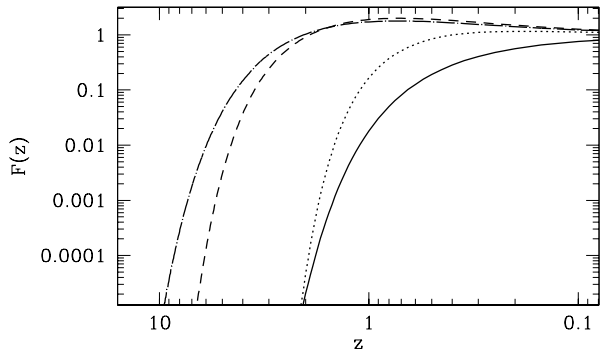}} \caption{The predicted
fractional rate of cluster formation as a function of redshift for
$\sigma_{8}\simeq 0.80$. The lines represent: (a) LSS$_{1}$ [dashed,
$(\beta,\gamma)=(-0.15,0.86)$], (b) LSS$_{2}$ [dot-dashed,
$(\beta,\gamma)=(0,0.66)$], (c) LJO (dot) and (d) $\Lambda$
cosmology (solid). Note that the clusters in LSS type models form
earlier ($z\sim 7-10$) in comparison to those produced in the
framework of the LJO and $\Lambda$CDM models ($z\sim 2$).}
\end{figure}

It is worth noticing that the above generic form of Eq.
(\ref{eq:89}) is heavily dependent on the choice of the background
cosmology and the matter power spectrum normalization, $\sigma_{8}$.
For the sake of comparison, in what follows we consider two values
$\sigma_{8}$. The first is the WMAP5 result $\sigma_{8} \simeq 0.80$
\cite{komatsu08} whereas the second is $\sigma_{8} \simeq 0.95$.

Finally, we normalize the above probability to give the fraction of
mass in bound structures which have already collapsed by the epoch
$z$, divided by the corresponding fraction in structures which have
collapsed at the present epoch ($z=0$), \be \tilde{F}(z)={\cal
F}(z)/{\cal F}(0) \;.\ee

In Fig. 2, by assuming the matter power spectrum normalization,
$\sigma_{8} \simeq 0.80$, we present in a logarithmic scale the
behavior of normalized structure formation rate as a function of
redshift for the present cosmological models.

\begin{figure}[ht]
\mbox{\epsfxsize=10cm \epsffile{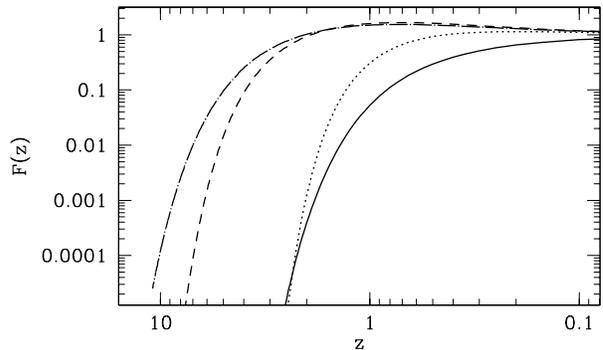}} \caption{The predicted
fractional rate of cluster formation as a function of redshift for
using a higher value for the normalization of the power spectrum,
$\sigma_{8}\simeq 0.95$. The meaning of the various lines is the
same of  Fig.\,2. We see that for a larger value of $\sigma_{8}$ the
corresponding cluster formation rate is displace to higher redshifts
(compare to Fig. 2).}
\end{figure}

In the context of the LSS$_{1}$ matter creation model (see dashed
line), we find that prior to $z\sim 0.8$ (or $z\sim 1.1$ for the
LSS$_2$ model, dot-dashed line) the cluster formation has
effectively terminated due to the fact that the matter fluctuation
field, $D(z)$, effectively overs (see section 5.3). Also, the large
amplitude of the LSS fluctuation field (see also Fig.\,1) implies
that in this model galaxy clusters appear to form much earlier
($z\sim 7$) with respect to the LJO (dot-dashed) and $\Lambda$
(solid line) cosmological models. Indeed, for the latter
cosmological models we find that galaxy clusters formed typically at
$z\sim 2.2$. From the observational point of view, it is interesting
to point here that the most distant cluster has been found at $z\sim
1.62$ \cite{Pap10}. We would like to stress that the LSS$_{2}$
predicts an early rapid cluster formation that takes place at the
redshift of reionization ($z\simeq z_{reion} \sim 10$), at which the
universe was reionized from the neutral state to a fully ionized
state. Obviously, this result rules out the LSS$_{2}$ model.

We also stress that the amount of cluster abundances between the two
models remains almost the same for $1.6 \le z \le 2.2$. This is to be
expected due to the fact that the two cosmological models have
almost the same growth factors (see also section 5.3). Then for
$z<1.6$ the LJO cluster formation rate becomes larger than that of
the concordance model and finally it terminates for $z \le 0.4$.

In Fig. 3  we also display the normalized structure formation rate
as a function of redshift, but now for a higher value of the power
spectrum parameter ($\sigma_{8}=0.95$). Comparing Figs. 2 and 3, we
see that the basic effect of a larger value of $\sigma_{8}$ is that
the corresponding cluster formation rate moves to higher redshifts.

\section{Conclusions}

In this work, we have investigated (analytically and numerically)
the overall dynamics of two cosmological models in which the dark
matter creation process provides  the late time accelerating phase
of the cosmic expansion without the need of dark energy. Such
scenarios termed here by LSS \cite{LSS08} and LJO \cite{LJO09} are
phenomenologically characterized by two distinct creation rates,
$\Gamma$ (see section 3).  In the first scenario (LSS), the creation
rate is $\Gamma = 3\gamma H_{0} + 3\beta H$ while in the second one
it is proportional to the Hubble parameter, namely:
$\Gamma=3\tilde{\Omega}_{\Lambda}(\rho_{co}/{\rho_{dm}})H$, where
$\rho_{co}=3H^{2}_{0}/8\pi G$  is the present day value of the
critical density.

It should also be stressed  that the phenomenological approach
adopted here
cannot determine the mass of the particles, as well as whether their
nature is fermionic or bosonic. In order to access the mass of the particles, and, therefore, the
nature of the CDM particles,  it is necessary to
determine the creation rate, $\Gamma$, from quantum field theory in FLRW
space time. In principle, such a treatment must somewhat  incorporate a
source of entropy since the matter creation process is truly an
irreversible process. In particular, in the case of adiabatic matter creation considered here, both the
entropy (S) and the number of particles (N) in a comoving volume
increase but the specific entropy (S/N) remains constant \cite{LCW,LG92,SLC02}.

In this context, by using current observational data (BAOs, CMB
shift parameter and SNIa) we have first performed a joint likelihood
analysis in order to put tight constraints on the main cosmological
parameters. Subsequently, trough a linear analysis we have also
studied the growth factor of density perturbations for both classes
of creation cold dark matter models, and, finally, by using the
Press-Schecther formalism we have discussed how the cluster
formation rates evolve in such scenarios.

The result of our joint statistical analysis shown that the fit
provided by the LSS model with $\gamma=0.66$ and $\beta=0$  turns
out to be of poor quality because it is unable to adjust
simultaneously the observational data at low and high redshift. This
result confirm the conclusion  by Steigman et al \cite{SSL09} using
only SNe Ia and the determination of $z_{eq}$, the redshift of the
epoch of matter radiation equality. For the LSS scenario, we also
find that the amplitude and the shape of the linear density contrast
are significantly different with respect to those predicted by the
LJO and $\Lambda$ models (see Fig. 1). In particular, for $z\le 0.8$
the matter fluctuation field of the LSS model practically decays
thereby implying that the corresponding cosmic structures cannot be
formed via gravitational instability (see also Figs. 2 and 3).

On the other hand, the creation cold dark matter scenario (termed
here as LJO) provides good quality fits of the cosmological
parameters at all redshifts and it resembles the global dynamics of
the concordance $\Lambda$CDM cosmology by including only one free
parameter. Despite, the latter the corresponding growth factor has
the following evolution with respect to that of the usual $\Lambda$
cosmology: (i) prior to the present epoch ($z<0.4$) the evolution of
the LJO growth factor reaches almost a plateau, which means that the
matter fluctuations are effectively frozen, (ii) between $0.4 \le z
< 1.6$, the growth factor in the LJO model is greater than that of
the concordance $\Lambda$ cosmology, while for $1.6 \le z \le 2.8$
the two growth factors have converged and (iii) for $z>2.8$ we find
that $D(z)<D_{\Lambda}(z)$ and the corresponding deviation
$(1-D/D_{\Lambda})\%$, remains close to $\sim 28\%$.

Summarizing, in the case of LSS scenario, the large scale structures
(such as galaxy clusters) form earlier ($z\sim 7-10$) with respect
to those produced in the framework of the LJO and $\Lambda$CDM
models ($z\sim 2$). Therefore, in view of the recent cluster
observational data, the latter are much more favored as compared to
the former. Our basic conclusion is that the LJO  creation cold dark
matter model which has only one free parameter passes all the tests
considered here. However, new constraints from complementary
observations  need to be investigated in order to see whether this
kind of scenario provides a realistic description of the observed
Universe.

\begin{acknowledgments}
SB and JASL would like to thank Manolis Plionis, Jos\'e
Fernando Jesus, Ribamar Reis and Gary Steigman,  for helpful
discussions. JASL is partially supported by CNPq and FAPESP under
grants 304792/2003-9 and 04/13668-0, respectively.
\end{acknowledgments}



\begin{thebibliography}{plain}

\bibitem {Spergel07}
 D.N. Spergel, et al., Astrophys. J. Suplem., {\bf 170}, 377 (2007).


\bibitem{essence}
T.M. Davis {\it et al.}, Astrophys. J., {\bf 666}, 716 (2007).

\bibitem {komatsu08}
E. Komatsu, et al., Astrophys. J. Suplem., {\bf 180}, 330 (2009);

\bibitem{Teg04}
M. Tegmark, et al., Astrophys. J., {\bf 606}, 702 (2004).

\bibitem{Eis05}
D. J. Eisenstein D. J., et al., Astrophys. J., {\bf 633}, 560
(2005); N. Padmanabhan, et al., Mon. Not. Roy. Soc., {\bf 378},
852(2007).

\bibitem{Kowal08}
M. Kowalski, et al., Astrophys. J., {\bf 686}, 749 (2008).

\bibitem{Hic09}
M. Hicken et al., Astroplys. J., {\bf 700}, 1097 (2009).

\bibitem {Weinberg89}
 S. Weinberg, Rev. Mod. Phys., {\bf 61}, 1 (1989).

\bibitem {Peebles03}
P. J. Peebles and B. Ratra, Rev. Mod. Phys., {\bf 75},
 559 (2003).

\bibitem{Pad03} T. Padmanabhan, Phys. Rept., {\bf 380}, 235 (2003).

\bibitem {Peri08}
L. Perivolaropoulos, [arxXiv.0811.4684], (2008).

\bibitem{coincidence} P.J.~Steinhardt,  in: \textit{Critical Problems in Physics}, edited by V.L. Fitch,
D.R. Marlow and M.A.E. Dementi (Princeton Univ. Pr.,  Princeton,
1997); P.J. Steinhardt, {Phil. Trans. Roy. Soc. Lond.} {\bf A361},
2497 (2003).

\bibitem{Reviews} J. A. S. Lima, Braz. Journ. Phys., {\bf 34}, 194
(2004), [astro-ph/0402109]; E. J. Copeland, M. Sami and S.
Tsujikawa, Int. J. Mod. Phys. {\bf{D15}}, 1753 (2006); M. S. Turner
and D. Huterer, Ann.~Rev.~Astron. \& Astrophys., {\bf 46}, 385
(2008).

\bibitem {Ratra88}
B. Ratra, and P. J. Peebles, Phys. Rev D., {\bf 37}, 3406 (1988).

\bibitem{Oze87}
M. Ozer M. and O. Taha, Nucl. Phys., {\bf B287}, 776, (1987); O. K.
Freese K., et al., Nucl. Phys., {\bf 287}, 797, (1987); O.
Bertolami, Nuovo Cimento B, {\bf 93B}, 36, (1986).

\bibitem{Free187} W. Chen and Y-S. Wu, Phys. Rev. {\bf D41}, 695 (1990); J. C. Carvalho, J. A. S. Lima, and I.
Waga, Phys. Rev. {\bf D46}, 2404 (1992); I. Waga, Astrophys. J. 414,
436 (1993);  J. A. S. Lima and J. M. F. Maia, Phys. Rev. D49, 5597
(1994); J. M. Overduin and F. I. Cooperstock, Phys. Rev. D., {\bf
58}, 043506, (1998); O. Bertolami and P. J. Martins, Phys. Rev. D.,
{\bf 61}, 064007 (2000); ; R. Opher and A. Pelinson, Phys. Rev. D.,
{\bf 70}, 063529 (2004); J. D. Barrow and T. Clifton, T., Phys. Rev.
D., {\bf 73}, 103520 (2006).

\bibitem{Wetterich:1994bg}
C.~Wetterich, Astron. Astrophys. {\bf 301}, 321 (1995)

\bibitem{Caldwell98}
R. R. Caldwell, R. Dave, and P. J. Steinhardt, Phys. Rev. Lett.,
{\bf 80}, 1582 (1998).

\bibitem{Brax:1999gp}
P.~Brax, and J.~Martin, Phys. Lett. {\bf B468}, 40 (1999).


\bibitem{KAM}
A. Kamenshchik, U. Moschella, and V. Pasquier, Phys. Lett. B., {\bf
511}, 265 (2001).

\bibitem{fein02}
A. Feinstein, Phys. Rev. D., {\bf 66}, 063511 (2002).

\bibitem{Caldwell}
R.R., Caldwell,  E. V., Phys. Rev. Lett., {\bf 545}, 23 (2002).

\bibitem{chime04}
L. P. Chimento, and A. Feinstein, Mod. Phys. Lett. A, {\bf 19}, 761
(2004).

\bibitem{Brookfield:2005td}
A. W. Brookfield, C.~van~de Bruck, D.~F. Mota, and
D.~Tocchini-Valentini, Phys. Rev. Lett. {\bf 96}, 061301 (2006).

\bibitem{Bauer05} F.~Bauer, Class.\ Quant.\ Grav.\  {\bf 22} (2005) 3533;
F. Bauer, Ph.d. Thesis, hep-th/0610178 (2006).

\bibitem{Grande06}
J. Grande, J. Sol\`a and H. \v{S}tefan\v{c}i\'{c}, JCAP {\bf 08},
{(2006)}, {011}; Phys. Lett. {\bf B645}, {236} (2007).



\bibitem{Boehmer:2007qa}
C.~G. Boehmer, T.~Harko, Eur. Phys. J. {\bf C50}, 423 (2007).


\bibitem{Bas09b}
S. Basilakos, Astron. Astrophys. {\bf 508}, 575 (2009); S.
Basilakos, M. Plionis, and J. Sol\'a, Phys. Rev. D., {\bf 80}, 3511
(2009).

\bibitem{FR} G. Allemandi, A. Borowiec, M. Francaviglia and
S. D. Odintsov, Phys. Rev. {\bf D 72}, 063505 (2005);
L. Amendola, D. Polarski and S. Tsujikawa, Phys. Rev. Lett. {\bf 98}, 131302 (2007);
J. Santos, J. S. Alcaniz, F. C. Carvalho and  N. Pires, Phys. Lett. {\bf B 669}, 14 (2008);
J. Santos and  M. J. Reboucas, Phys. Rev. D {\bf 80}, 063009 (2009).
For a review see, T. P. Sotiriou and V. Faraoni, arXiv:0805.1726v2, [gr-qc], Rev. Mod. Phys. (2009).

\bibitem{Nat10} R. Reyes {\it et al.}, Nature {\bf 464}, 256 (2010).

\bibitem{Inhom} T. Buchert,  Gen. Rel. Grav 9 306 (2000); P. Apostolopoulos {\it et al.}  JCAP P06 009 (2006); T.  Kai {\it et al.}, Prog. Theor. Phys. {\bf 117}, 229 (2007); E. W.  Kolb, S.  Matarrese, A.  Notari  and A. Riotto, Phys Rev {\bf D71}, 023524 (2005);  R. Sussman, arXiv:1002.0173v1 [gr-qc] (2010).

\bibitem{Quart} N. Bilic, G. B. Tupper, and R. D. Violler, Phys. Lett. B535, 17
(2002); M. C. Bento, O. Bertolami and A. A. Sen, Phys. Rev. D66,
043507 (2002); M. Makler, S. Q. de Oliveira, and I. Waga, Phys.
Lett. B555, 1 (2003); J. C. Fabris, S. V. B. Goncalves, and P. E. de
Souza, [astroph/ 0207430]; J. V. Cunha, J. S. Alcaniz and J. A. S.
Lima, Phys. Rev. D {\bf 69}, 083501 (2004) [astro-ph/0306319];  B.
Wang, C. Y. Lin, and E. Abdalla, Phys. Lett. {\bf B637}, 357 (2006);
J. A. S. Lima, J. V. Cunha and J. S. Alcaniz, Astropart. Phys. {\bf
31}, 233 (2009), astro-ph/0611007; {\bf ibdem}, Astropart. Phys.
{\bf 30}, 196 (2008).

\bibitem{SCHO39} E. Schr\"odinger, Physica VI, 899 (1939).

\bibitem{Parker} L. Parker, { Phys. Rev. Lett.} {\bf 21}, 562 (1968);
{\it Phys. Rev.} {\bf 183}, 1057 (1969); S. A. Fulling, L. Parker
and B. L. Hu, {Phys. Rev.} {\bf 10}, 3905 (1974).


\bibitem{BirrellD} N. D. Birrell and P. C. Davies, {\it Quantum Fields in
Curved Space}, Cambridge Univ. Press, Cambridge, (1982); A. A. Grib,
S. G. Mamayev and V. M.  Mostepanenko,{\it  Vaccum Quantum Effects
in Strong Fields},  Friedmann Laboratory Publishing, St. Petersburg
(1994);

\bibitem{Parker1} V. K. Onemli and R. P. Woodard, Phys. Rev. D {\bf 70}, 107301 (2004); L. Parker and A. Raval, Phys. Rev. Lett. {\bf 86}, 749 (2001).

\bibitem{MW07} V. F. Mukhanov and S. Winitzki, {\it Introduction to Quantum
Effects in Gravity}, Cambridge UP, Cambridge (2007).

\bibitem{Prigogine} I. Prigogine et al., Gen. Rel. Grav., {\bf
21}, 767 (1989).

\bibitem{LCW} J. A. S. Lima, M. O. Calv\~{a}o, and I. Waga, ``Cosmology,
Thermodynamics and Matter Creation'', {\it Frontier Physics, Essays
in Honor of Jayme Tiomno}, World Scientific, Singapore (1990),
[arXiv:0708.3397]; M. O. Calv\~{a}o, J. A. S. Lima, and I. Waga,
Phys. Lett. {\bf A162}, 223 (1992).

\bibitem{LG92} J. A. S. Lima, A. S. M. Germano, Phys. Lett. A {\bf 170},
373 (1992).

\bibitem{ZP2} W. Zimdahl and D. Pav\'on, Phys. Lett. A {\bf 176},
57 (1993); R. A. Susmann, Class. Q. Grav. {\bf 11}, 1445 (1994); W.
Zimdahl and D. Pav\'{o}n, Mon. Not. R. Astr. Soc. {\bf 266}, 872
(1994); W. Zimdahl and D. Pav\'{o}n, GRG {\bf 26}, 1259  (1994); J.
Gariel and G. Le Denmat, Phys. Lett. A {\bf 200}, 11 (1995).

\bibitem{LGA96} J. A. S. Lima, A. S. M. Germano and
L. R. W. Abramo, Phys. Rev. D {\bf 53}, 4287 (1996); L. R. W. Abramo
and J. A. S. Lima, Class. Quant. Grav. {\bf 13}, 2953 (1996); J. A.
S. Lima and J. S. Alcaniz,  Astron. Astrophys. {\bf 348}, 1 (1999);
W. Zimdahl, D. J. Schwarz, A. B. Balakin  and D. Pav\'on, Phys. Rev.
D {\bf 64}, 063501 (2001); M. de Campos and J. A. Souza, Astron.
Astrophys. {\bf 422}, 401 (2004); Y. Quiang, T-J. Zhang and Z-L. Yi,
Astrop. Space Sci. {\bf 311}, 407 (2007).

\bibitem{LA99} J. A. S. Lima and J. S. Alcaniz, Astron.
Astrophys. {\bf 348}, 1 (1999); J. S. Alcaniz and J. A. S. Lima,
Astron. Astrophys. {\bf 349}, 729 (1999).

\bibitem{Susmann94}  R. A. Susmann, Class. Q. Grav. {\bf 11}, 1445 (1994);
J. Gariel, and M. Cissoko, Class. Quant. Grav. {\bf 13}, 2977
(1996); L. R. W. Abramo and J. A. S. Lima, Class. Quantum Grav. {\bf
13}, 2953 (1996); V. B. Johri and S. K. Pandey, Int. J. Theor. Phys.
{\bf 38}, 1981 (1999); T. Harko and M. K. Mak, Astrophys. Space Sci.
{\bf 253}, 161 (1997); J. A. S. Lima, Gen. Rel. Grav. {\bf 29}, 805
(1997); K. Desinkhan, Gen. Rel. Grav. {\bf 29}, 435 (1997); Class.
Quant. Grav. {\bf 16}, 2741 (1999); J. A. S. Lima and  L. R. W.
Abramo, Phys. Lett A, {\bf 257}, 123 (1999); M. K. Mak and T. Harko,
Class. Quant. Grav. {\bf 16}, 4085 (1999); {\bf ibdem}, IJMP D {\bf
8}  607 (1999); J. A. S. Lima and U. F. Wichoski, Phys. Lett. A {\bf
262}, 103 (1999); W. Zimdahl, Phys. Rev. D {\bf 61} 083511 (2000);
J. A. S. Lima, A. I. Silva and S. M. Viegas, Mon. Not. R. Astron.
Soc. {\bf 312}, 747 (2000); G. Montanni, Class. Quant. Grav. {\bf
18}, 193 (2001);  G. P. Singh, R. V. Deshpande, T. Singh, Astrop.
Space Sci. {\bf 282}, 489 (2002); J. A. S. Lima and J. A. E.
Carrilo, [astro-ph/0201168] (2002); M. de Campos, Gen. Rel. Grav.
{\bf 35}, 899 (2003); M. de Campos and J. A. Souza, Astron.
Astrophys. {\bf 422}, 401 (2004).

\bibitem{ZSBP01} W. Zimdhal, D. J. Schwarz, A. B. Balakin  and D. Pav\'on, Phys. Rev. D {\bf 64}, 063501
(2001).


\bibitem{SLC02} R. Silva, J. A. S. Lima and M. O. Calv\~ao, Gen. Rel. Grav. {\bf 34},
865 (2002).

\bibitem{freaza02} M. P. Freaza, R. S. de Souza and I. Waga, Phys. Rev. D {\bf 66}, 103502 (2002).

\bibitem{Makler07} C. S. Camara, J. C. Carvalho and M. R. G. Maia,  IJMP D {\bf 16}, 427 (2007);
Y. Quiang, T-J. Zhang and Z-L. Yi, Astrop. Space Sci. {\bf 311}, 407
(2007); S. S. Costa and M.  Makler, [astro-ph/0702418] (2007).

\bibitem{LSS08} J. A. S. Lima, F. E. Silva and R. C. Santos, Class. Quant. Grav. {\bf 25}, 205006
(2008), arXiv:0807.3379 [astro-ph]

\bibitem{SSL09} G. Steigman, R. C. Santos and J. A. S. Lima, JCAP {\bf
0906}, 033 (2009), arXiv:0812.3912 [astro-ph]


\bibitem{LJO09}
J. A. S. Lima, J. F. Jesus and F. A. Oliveira, arXiv:09115727
(2009).

\bibitem{Bond:1997wr}
  J.~R.~Bond, G.~Efstathiou and M.~Tegmark,
  MNRAS  {\bf 291}, L33 (1997).

\bibitem{Nesseris:2006er}
  S.~Nesseris and  L.~Perivolaropoulos,
  JCAP {\bf 0701}, 018 (2007).

\bibitem{Elgaroy07} O. Elgaroy \& T. Multamaki, JCAP {\bf 9}, 2 (2007);
 P.S. Corasaniti \& A. Melchiorri Phys. Rev. D {\bf 77}, 103507 (2008).

\bibitem{LZB97} J. A. S. Lima, V. Zanchin and R.  Brandenberger,  MNRAS {\bf 291}, L1 (1997).

\bibitem{Reis03} R. R. R. Reis, Phys. Rev. D {\bf 67}, 087301 (2003).

\bibitem{FZ} J. C. Fabris, S.V. B. Gon\c{c}alves, H. E. S. Velten, and W. Zimdahl, Phys. Rev. D 78, 103523 (2008).

\bibitem{Pee93} P. J. E. Peebles, Principles of Physical Cosmology, Princeton
University Press, Princeton New Jersey (1993).


\bibitem{Arc94} R. C. Arcuri and I. Waga., Phys. Rev. D. {\bf 50}, 2928 (1994).

\bibitem{Pap10} C. Papovich, et al., [arXiv:1002.3158] (2010).

\bibitem{press}  W.~H.~Press and P.~Schechter, Astrophys. J. {\bf 187}, 425 (1974).

\bibitem{Peeb84} P. J. E. Peebles, Astrophys. J. {\bf 284}, 439, (1984); S. Weinberg,
{\bf 59}, 2607 (1987).

\bibitem{Rich92}
D. Richstone, A. Loeb and E. L. Turner, Astrophys. J. {\bf 393}, 477
(1992)

\bibitem{LaceyCole}
C. Lacey and S. Cole, Mon. Not. R.
Astr. Soc, {\bf 262},627 (1993).

\bibitem{eke}
V. Eke, S. Cole and C. S. Frenk, MNRAS {\bf 282}, 263 (1996).

\bibitem{Bas1}
S. Basilakos, Astrophys. J. {\bf 590}, 636 (2003).

\bibitem{Bas2}
S. Basilakos, J.C.B. Sanchez, and L. Perivolaropoulos, Phys. Rev. D,
{\bf 80}, 043530 (2009).

\bibitem{Bas3}
S. Basilakos, M. Plionis, and J. Sol\`a, Phys.Rev.D {\bf 80}, 083511
(2009).



\end{thebibliography}
\end{document}